\documentclass[%
 reprint,
 amsmath,amssymb,
 aps,
]{revtex4-2}

\usepackage{graphicx}
\usepackage{dcolumn}
\usepackage{bm}

\usepackage{graphicx}
\usepackage{dcolumn}
\usepackage{bm}
\usepackage{upgreek} 
\usepackage{graphicx}    
\usepackage{ragged2e}  
\usepackage[percent]{overpic}
\usepackage[T1]{fontenc}
\usepackage[utf8]{inputenc}

\begin{document}

\preprint{APS/123-QED}

\title{Nematic Fluctuations and Electronic Correlations in Heavily Hole-Doped Ba$_{1-x}$K$_x$Fe$_2$As$_2$ Probed by Elastoresistance}

\author{Franz Eckelt}
\email{eckelt@uni-wuppertal.de}
\affiliation{University of Wuppertal, School of Mathematics and Natural Sciences, 42097 Wuppertal, Germany}

\author{Steffen Sykora}
\affiliation{Leibniz-Institute for Solid State and Materials Research, 01069 Dresden, Germany}

\author{Xiaochen Hong}
\affiliation{University of Wuppertal, School of Mathematics and Natural Sciences, 42097 Wuppertal, Germany}
\affiliation{Department of Applied Physics and Center of Quantum Materials and Devices, Chongqing University, 401331 Chongqing, China}

\author{Vilmos Kocsis}
\affiliation{Leibniz-Institute for Solid State and Materials Research, 01069 Dresden, Germany}

\author{Vadim Grinenko}
\affiliation{Tsung-Dao Lee Institute, Shanghai Jiao Tong University, Shanghai 201210, China}

\author{Bernd B\"uchner}
\affiliation{Leibniz-Institute for Solid State and Materials Research, 01069 Dresden, Germany}

\author{Kunihiro Kihou}
\affiliation{National Institute of Advanced Industrial Science and Technology, Tsukuba, Ibaraki 305-8568, Japan}

\author{Chul-Ho Lee}
\affiliation{National Institute of Advanced Industrial Science and Technology, Tsukuba, Ibaraki 305-8568, Japan}

\author{Christian Hess}
\email{c.hess@uni-wuppertal.de}
\affiliation{University of Wuppertal, School of Mathematics and Natural Sciences, 42097 Wuppertal, Germany}


\date{\today} 

\begin{abstract}
This work investigates nematic fluctuations and electronic correlations in the hole-doped iron pnictide superconductor Ba$_{1-x}$K$_x$Fe$_2$As$_2$ by means of longitudinal and transverse elastoresistance measurements over a wide doping range ($0.63 < x < 0.98$). For this purpose, the orbital character of the electronic response was revealed by decomposition of the elastoresistance into the $A_{1g}$ and $B_{2g}$ symmetry channels. It was shown that at lower doping levels nematic fluctuations in the $B_{2g}$ channel dominate, while for $x > 0.68$ the $A_{1g}$ channel becomes dominant and reaches a pronounced maximum at $x \approx 0.8$ which indicates strong orbital-selective electronic correlations. Despite the dominance of the $A_{1g}$ signal at high doping, a weak contribution in the $B_{2g}$ channel persists, which can be interpreted as a remnant of nematic fluctuations. Model calculations based on a five-orbital tight-binding Hamiltonian with interactions attribute the observed enhancement in the $A_{1g}$ channel to an orbital-selective Kondo-like resonance, predominantly involving the $d_{xy}$ orbital. We discuss our results in relation to the evolution of the Sommerfeld coefficient reported in the literature and a reported change of the superconducting order parameter. All this indicates that for $x > 0.68$ qualitatively new physics emerges. Our findings suggest that electronic correlations in the strongly hole-doped regime play an important role in superconductivity, while the detectable weak nematic fluctuations may also be of relevance.
\end{abstract}

\maketitle


\section{Introduction}
Nematicity, the spontaneous breaking of rotational symmetry in the electronic system, is a ubiquitous phenomenon in iron-based superconductors and is considered a key ingredient in understanding their unconventional superconductivity~\cite{Fernandes2014,Gallais2013}. In hole- and electron-doped variants of BaFe$_2$As$_2$, such as Ba$_{1-x}$K$_x$Fe$_2$As$_2$, nematic fluctuations dominate the low-doping region of the phase diagram, particularly in the $B_{2g}$ symmetry channel of the $D_{4h}$ point group~\cite{Bohmer2014}. These fluctuations, often linked to anisotropic spin and orbital interactions, have been proposed to enhance superconducting pairing~\cite{Eckberg2020}.
Elastoresistance has emerged as a sensitive probe to disentangle the symmetry-resolved electronic response to external strain. In the tetragonal $D_{4h}$ point group, the structural distortion associated with nematic order transforms according to the $B_{2g}$ representation, which corresponds to a strain mode that breaks the fourfold rotational symmetry by elongating one Fe--Fe bond while compressing the orthogonal one \cite{Wieki2020}. Nematic fluctuations therefore predominantly couple to antisymmetric $B_{2g}$-type strains, allowing their detection via elastoresistance measurements in the corresponding symmetry channel.\\
At higher hole doping, the electronic landscape of Ba$_{1-x}$K$_x$Fe$_2$As$_2$ undergoes profound changes. As the system approaches a nominal Fe 3d$^5$ configuration, correlation effects become increasingly prominent, leading to orbital-selective behavior~\cite{Hardylang,Misawa2012}. The concept of orbital-selective Mottness has emerged as a key framework to understand the electronic properties of hole-doped iron-based superconductors~\cite{TheorieMott}. It is now well established that the strong Hund’s coupling in these systems decouples charge excitations between different Fe 3$d$ orbitals, causing each orbital to behave as an effectively independent, doped Mott insulator. As a result, the degree of electronic correlation becomes strongly orbital dependent and increases with hole doping, especially in the $d_{xy}$ orbital, which approaches half-filling more rapidly than others. This framework provides a microscopic foundation for the coexistence of weakly and strongly correlated electrons and supports the observed doping-dependent crossover from nematic fluctuation-dominated behavior at low doping to correlation-driven physics at high doping in Ba$_{1-x}$K$_x$Fe$_2$As$_2$.
In this high-doping regime, the electronic correlations are expected to manifest in the elastoresistive response primarily via the $A_{1g}$ channel, which corresponds to in-plane symmetric strain. Consistent with this picture, elastoresistance studies on the alkali end members of the 122 family KFe$_2$As$_2$, RbFe$_2$As$_2$, and CsFe$_2$As$_2$ have revealed a pronounced $A_{1g}$ response and a strongly suppressed $B_{2g}$ signal~\cite{Wiecki2021}. This result underscores the dominance of electronic correlations over nematic fluctuations in the strongly hole-doped regime.
Simultaneously, the superconducting gap structure in this regime appears to evolve significantly. Around $x \approx 0.7$--0.8, recent $\mu$SR and thermoelectric measurements revealed a superconducting phase with broken time-reversal symmetry (BTRS), suggesting a complex $s+is$ pairing symmetry driven by competing interband interactions and Fermi surface topology changes~\cite{Vadim2020,Grinenko2021neu,Shipulin2023}. These findings highlight the rich interplay of correlations, nematicity, and superconductivity in this system.
Previous elastoresistivity studies on heavily hole-doped Ba$_{1-x}$K$_x$Fe$_2$As$_2$ reported a pronounced maximum of the longitudinal elastoresistance around $x \approx 0.8$, which was attributed to a Lifshitz transition in the electronic structure~\cite{Xiaochen1}. While this interpretation was supported by earlier theoretical work, the role of nematic fluctuations and electronic correlations in this regime remains insufficiently understood. In this study, we revisit this doping region using both experimental data and refined theoretical calculations that include additional interaction effects, aiming to clarify the microscopic origin of the observed elastoresistance behavior. We present elastoresistance measurements on Ba$_{1-x}$K$_x$Fe$_2$As$_2$ single crystals across a broad doping range ($x = 0.6$--$0.95$), where both nematic fluctuations and correlation effects are known to evolve. We decompose the experimental results of the elastoresistive response into $B_{2g}$ and $A_{1g}$ channels. By a comparison of these results with theoretical calculations based on a realistic multiorbital Hubbard model, we identify a crossover from nematic-dominated behavior at low $x$ to a correlation-driven $A_{1g}$ response at high $x$. 

Our findings offer a unified picture of how distinct fluctuation channels shape the phase diagram of Ba$_{1-x}$K$_x$Fe$_2$As$_2$ and point toward the role of strain as a tuning knob to explore intertwined electronic orders in correlated superconductors.

\section{Methods}

\begin{figure}[!htbp]
\includegraphics[width=0.49\textwidth]{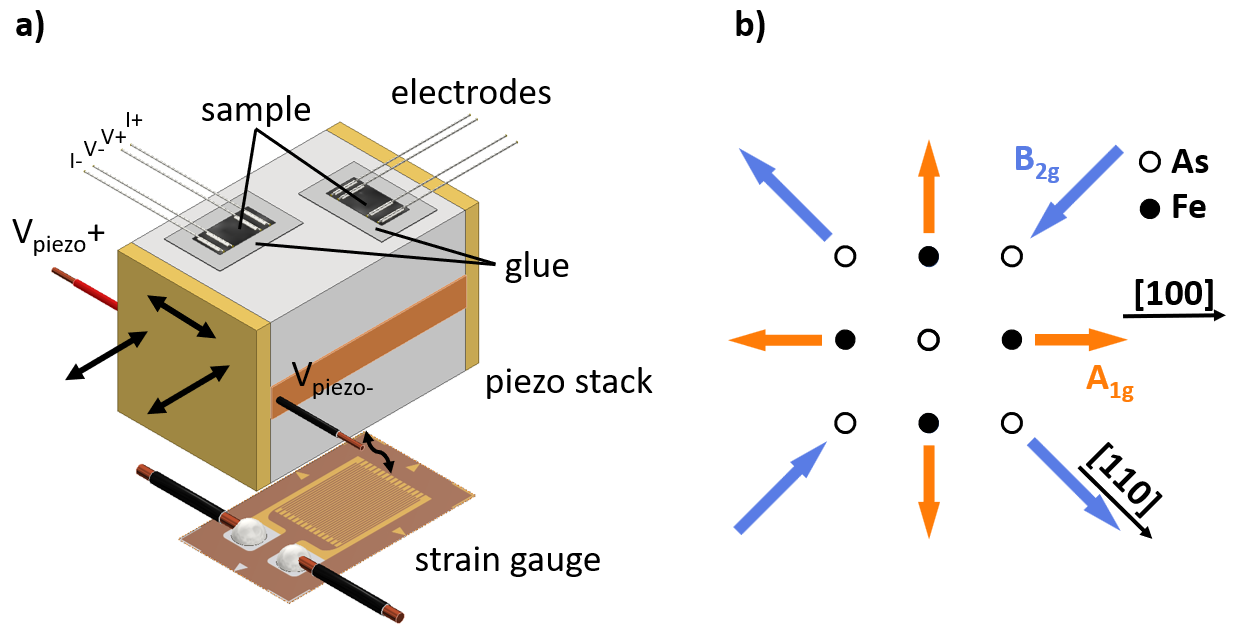}
\caption{(a) Schematic of the fully contacted piezoelectric actuator. (b) Schematic representation of the $A_{1g}$ and $B_{2g}$ symmetry channels of the $D_{4h}$ point group.}
\label{fig:Piezo}
\end{figure}

The heavily hole-doped Ba$_{1-x}$K$_x$Fe$_2$As$_2$ single crystals ($x$ =0.63,0.67,0.72,0.74,0.80,0.81,0.85,0.95,0.98 ) were grown using the self-flux method \cite{Kristallzüchtungfinal}. 
Elastoresistance measurements were performed following the procedures described in Refs. \cite{elektronen1,FeSeS,Xiaochen2020LaFe,Xiaochen1}. 
The samples were cut into thin bars with dimensions of approximately 
1\,mm\,$\times$\,0.5\,mm and a thickness of about 40\,$\upmu$m, and contacted 
with four 50\,$\upmu$m silver wires for four-point resistance measurements. 
For contacting, either a combination of Devcon No.~14250 and EPO-TEK H20E was used, 
or, for some samples, Hans Wolbring 200N. The EPO-TEK H20E was cured at 120\,$^{\circ}$C for 15\,min inside an argon-filled glovebox in order to prevent degradation of the potassium-containing samples due to oxygen or moisture. The prepared sample was glued onto a piezoelectric actuator (PSt150/5x5/7 from Piezomechanik GmbH), and the applied strain was measured using a strain gauge (N5K-06-S5030K-50C/DG/E4 from Micro-Measurements) that was also attached to the backside of the actuator. A schematic illustration of the fully prepared piezoelectric actuator is shown in Fig. \ref{fig:Piezo}\,a).
\\
The sample resistance was collected with a combination of a high-precision current
source (Keithley 2400 SourceMeter) and a nanovoltage meter (Keithley 2182 NanovoltMeter). Special care was taken to avoid a temperature drift effect, and the electric current was set in an alternating positive/negative manner to avoid artifact.

\section{Results}
\begin{figure*}[!htbp]
    \centering
    \includegraphics[width=0.98\textwidth]{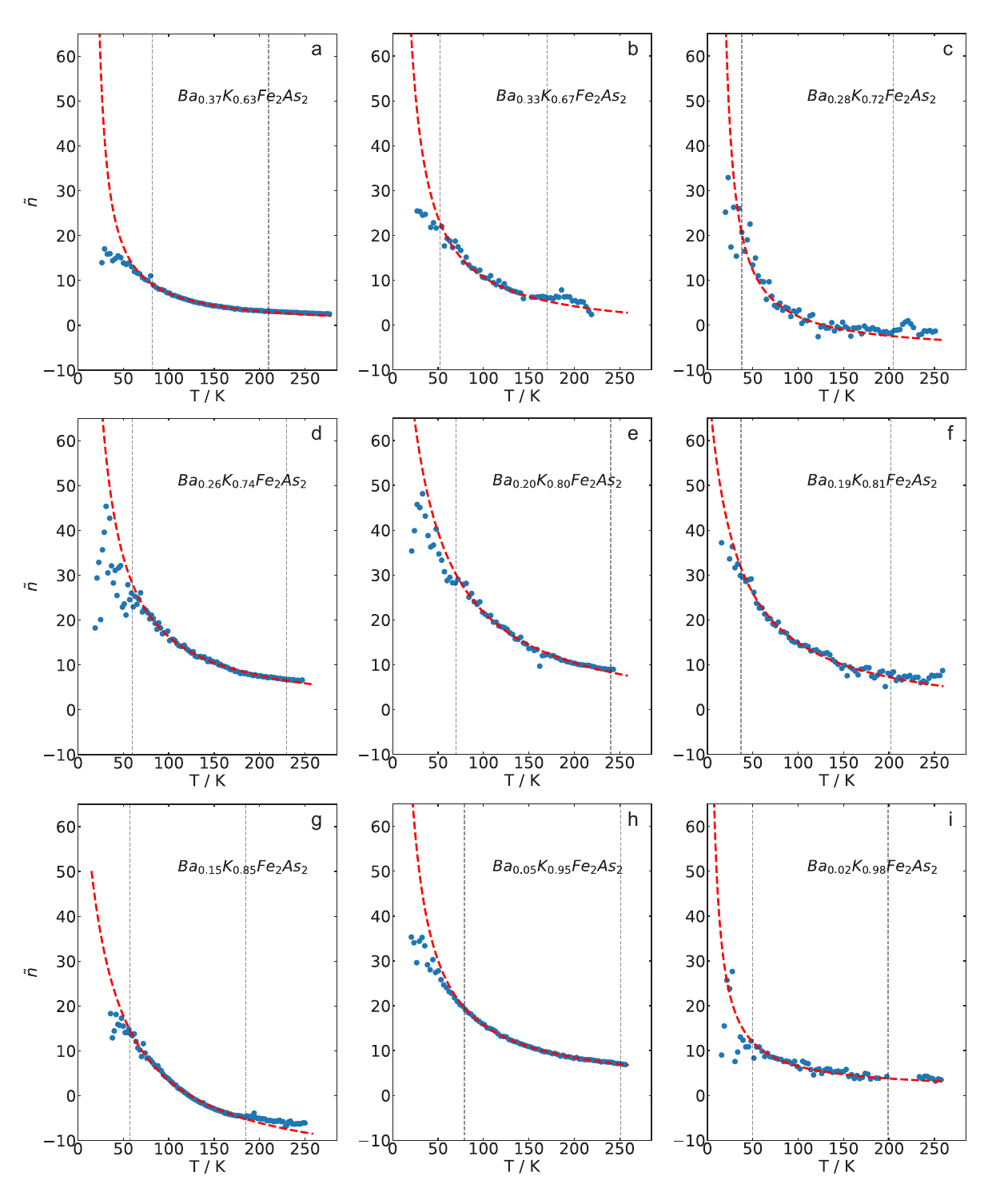}
    \caption{Longitudinal elastoresistance measurements of Ba$_{1-x}$K$_x$Fe$_2$As$_2$ samples with 0.63 $\leq  \times \leq$ 0.98 (blue) and the corresponding Curie-Weiss fits (red). The vertical lines indicate the fitting range for each sample. }
    \label{fig:Elastomessungen}
\end{figure*}

The longitudinal elastoresistance was measured for nine samples (see Fig. \ref{fig:Elastomessungen}). The observed temperature dependence follows a Curie-Weiss-type relation given by \cite{Xiaochen1}:

\begin{equation}
\tilde{n}=\tilde{n}_0 + \frac{a}{T-b}
\label{eq:Curie}
\end{equation}

\begin{figure*}[!htbp]
    \centering

    \begin{minipage}[t]{0.49\textwidth}
        \centering
        \begin{overpic}[height=5.5cm]{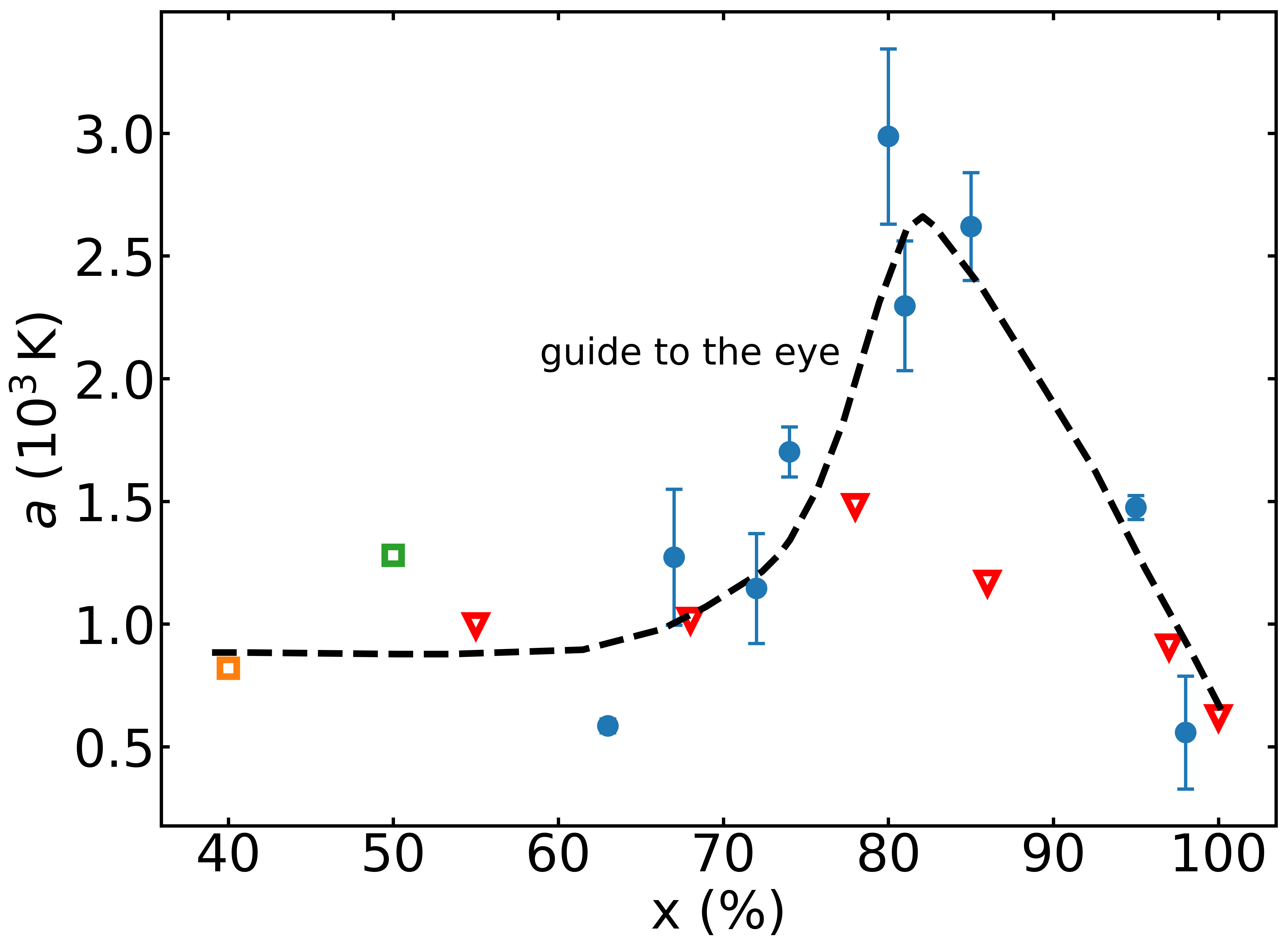}
            \put(2,75){\bfseries (a)}
        \end{overpic}
        \label{fig:Ergebnisse_a}
    \end{minipage}
    \hfill
    \begin{minipage}[t]{0.49\textwidth}
        \centering
        \begin{overpic}[height=5.5cm]{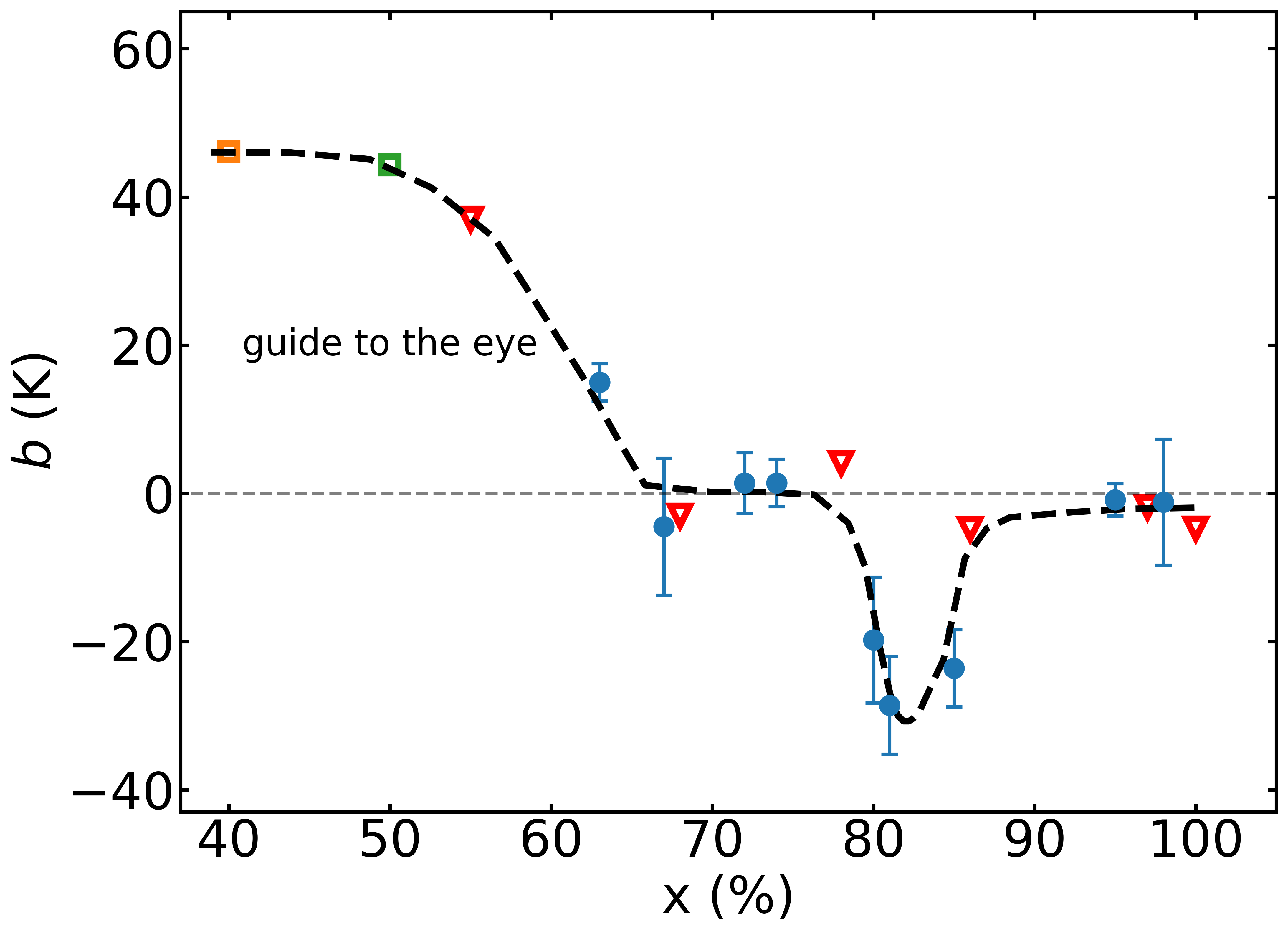}
            \put(2,75){\bfseries (b)}
        \end{overpic}
        \label{fig:Ergebnisse_b}
    \end{minipage}

    \caption{\justifying
    Summary of the Curie-Weiss fit parameters extracted from the longitudinal elastoresistance measurements shown in Fig.~\ref{fig:Elastomessungen} for the different doping levels. The dataset has been complemented with values from the literature. The orange rectangle is based on values from Ref.~\cite{Daten1}, the green rectangle on Ref.~\cite{Daten2}, and the red triangles correspond to the measurements reported in Ref.~\cite{Xiaochen1}. The connecting lines between the data points are guides to the eye.
    }
    \label{fig:Ergebnisse}
\end{figure*}

The performed Curie-Weiss fits are shown in Fig. \ref{fig:Elastomessungen} as red dashed lines. The fitting procedure follows the method described in Ref.~\cite{Daten1}, and the corresponding temperature range used for the fits is marked by vertical lines. The fit parameters $a$ and $b$ are presented in Fig. \ref{fig:Ergebnisse} as a function of doping. Additionally, the values are listed in Table \ref{tab:Fitparameter_longitudinal}. Literature values, indicated by open symbols, have been included for comparison.

\begin{table*}[!htbp]
\caption{Fit parameters of the Curie-Weiss fits to the longitudinal elastoresistance measurements. The errors reported correspond to the estimated standard deviation (1$\sigma$) of the respective parameter based on the covariance matrix of the fit.}
\centering
\small
\begin{ruledtabular}
\begin{tabular}{ccccc}
$x$  & $a\,$/\,K & $b\,$/\,K & $n_0$& Temperature range / K  \\ \hline
0.63  &   $585\pm 28$  & $15.7 \pm 2.5 $   & $0.1 \pm 0.1 $  & 82 - 210 \\ 
0.67  &  $1272 \pm 277 $   & $-4.5 \pm 9.2 $   & $-2.5 \pm 1.3 $& 52 - 170   \\ 
0.72  & $1145 \pm 224$    & $1.4 \pm 4.1 $   & $-6.2 \pm 0.8 $&38 - 205    \\ 
0.74  & $1702 \pm 102$    & $1.4 \pm  3.2$   & $0.9 \pm 0.4 $&70 - 230    \\ 
0.80  & $2987 \pm 357$    & $-19.8 \pm 8.5 $   & $-3.1 \pm 1.2 $ &70 - 240   \\
0.81  & $2297 \pm 264$    & $-28.6 \pm 6.6 $   & $2.6 \pm 1.1 $& 37 - 202   \\ 
0.85  & $2620 \pm 220$    & $-23.6 \pm  5.2$   & $-17.8 \pm 0.8 $&57 - 185    \\ 
0.95  & $1475 \pm 50$    & $-0.9 \pm  2.2$   & $1.12 \pm 0.2 $&79 - 251    \\ 
0.98  & $558 \pm 230$    & $-1.2 \pm  8.5$   & $-1.0 \pm 1.1 $& 50 - 199   \\ 
\end{tabular}
\label{tab:Fitparameter_longitudinal}
\end{ruledtabular}
\end{table*}

A pronounced maximum of the $a$ parameter is observed at a doping level of $x \approx 0.8$, consistent with the red-triangle data reported by Hong et al.~\cite{Xiaochen1}. In their study, the $b$ parameter decreases to $0\,\mathrm{K}$ up to $x \approx 0.67$. In the present work, however, a local minimum emerges at $x \approx 0.8$.

\begin{figure*}
    \centering
    \includegraphics[width=0.90\textwidth]{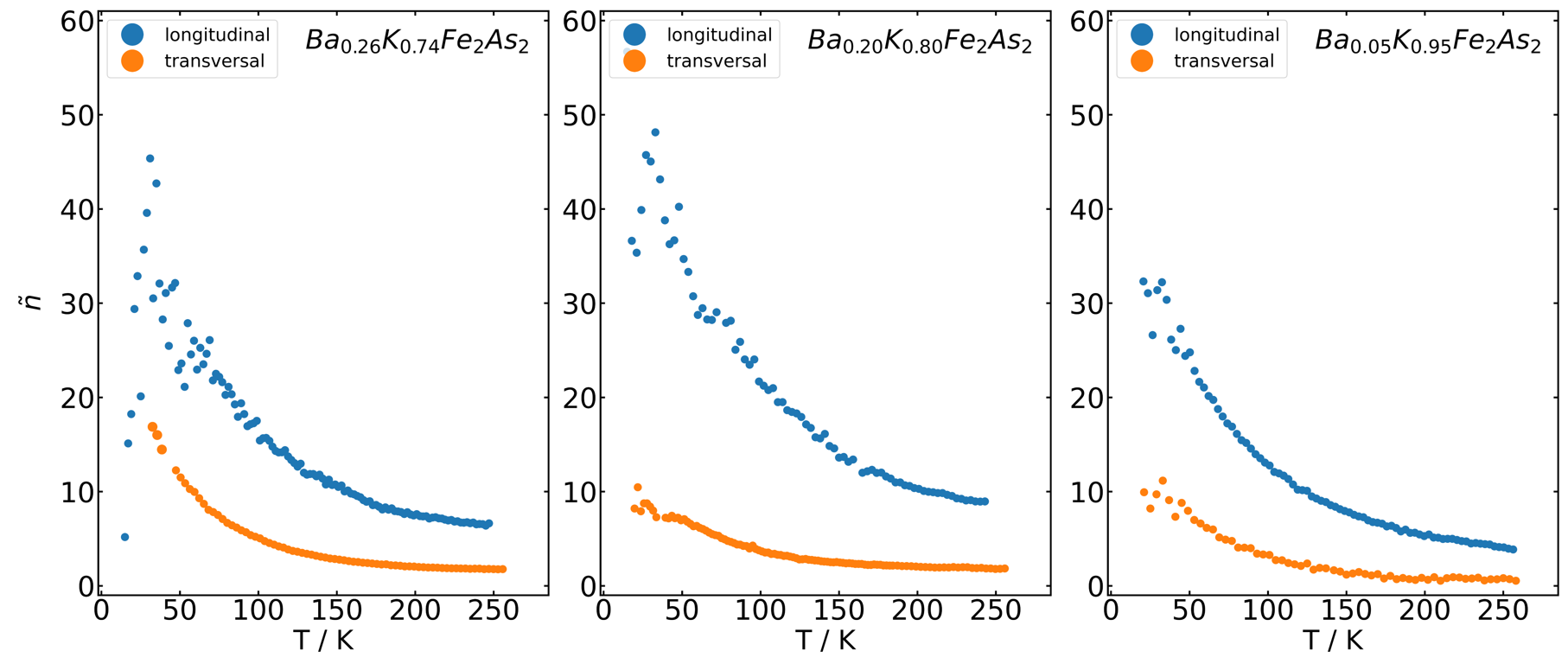}   
    \caption{The previously shown longitudinal elastoresistance measurements (blue) were complemented by the transverse resistance measurements shown in orange.  }
    \label{fig:Daten_beideRichtungen}
\end{figure*}

To investigate the increase of the $a$-parameter in more detail, the transverse elastoresistance was additionally measured for three samples with $x = 0.74$, $0.8$, and $0.95$ along the [110] direction. These doping levels correspond to the rising flank, the maximum, and the falling flank of the $a$-parameter. The longitudinal (blue) and transverse (orange) elastoresistance for these samples are shown in Fig. \ref{fig:Daten_beideRichtungen}.

As shown in Ref. \cite{Wiecki2021}, it is essential to decompose the divergent elastoresistance into its individual symmetry channel contributions in the strongly hole-doped compounds. Elastoresistance measurements along the [110] direction allow for the extraction of the $A_{1g}$ and $B_{2g}$ channel components as follows \cite{Wieki2020}:
\begin{equation} 
m_{A_{1g}}= \frac{1}{1-\nu}\Bigg[\frac{d(\Delta R /R_0)_{[110]}}{d\epsilon_{[110]}}+\frac{d(\Delta R /R_0)_{[\bar{1}10]}}{d\epsilon_{[110]}}\Bigg],
\label{eq:A1g} 
\end{equation} 
\begin{equation}
m_{B_{2g}}= \frac{1}{1+\nu}\Bigg[\frac{d(\Delta R /R_0)_{[110]}}{d\epsilon_{[110]}}-\frac{d(\Delta R /R_0)_{[\bar{1}10]}}{d\epsilon_{[110]}}\Bigg].
\label{eq:B2g}
\end{equation}
Here, $\epsilon_{[110]}$ denotes the applied strain along the $[110]$ crystallographic direction, $\Delta R$ is the strain-induced change of the resistance, $R_0$ is the resistance of the unstrained sample, and $\nu$ is the Poisson ratio of the piezoelectric actuator.

\begin{figure*}
    \centering
    \includegraphics[width=0.9\textwidth]{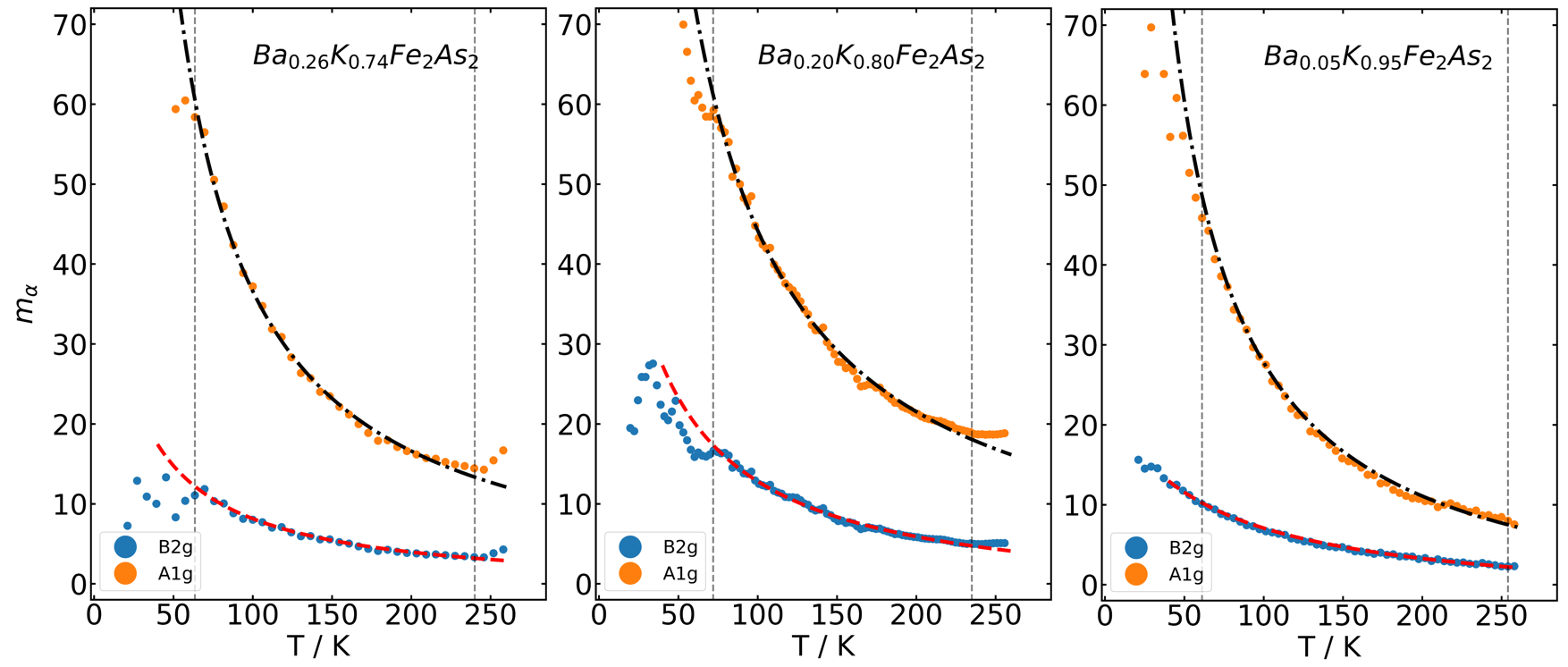}   
    \caption{ Results of the elastoresistance decomposition into the $A_{1g}$ and $B_{2g}$ symmetry channels, determined using Equation \ref{eq:A1g} and \ref{eq:B2g}, for samples with doping levels of 0.74, 0.80, and 0.95. All samples exhibit a divergent increase in both the $A_{1g}$ and $B_{2g}$ channels, which can also be described by a Curie-Weiss-like behavior. However, the signal in the $A_{1g}$ channel is significantly more pronounced than in the $B_{2g}$ channel for all samples. }
    \label{fig:Zerlegung}
\end{figure*}

This decomposition is shown in Fig. \ref{fig:Zerlegung}. For all three samples, both the $A_{1g}$ (orange) and $B_{2g}$ (blue) channels exhibit a divergent behavior with decreasing temperature, with the $A_{1g}$ response being significantly larger in magnitude than the $B_{2g}$ signal in each case. These data can also be described by Curie-Weiss-like behavior, and the corresponding fit parameters are listed in Table \ref{tab:Fitparameter}.

\begin{table*}
\caption{Fit parameters of the Curie--Weiss fits to the $A_{1g}$ and $B_{2g}$ channel signals. 
The reported errors correspond to the estimated standard deviation (1$\sigma$) of the respective parameter, based on the covariance matrix of the fit.}
\centering
\small
\begin{ruledtabular}
\begin{tabular}{cccccc}
 $x$ & Kanal & $a\,$/\,K & $b\,$/\,K & $\tilde{n}_0$& Temperature range / K  \\ \hline
0.74 & $A_{1g}$ & $3786\pm 250$  & $3.7\pm 3.3$ & $-2.7\pm 1.1$ &60 - 240   \\ 
     & $B_{2g}$ &$1060 \pm 147$  &$-17.3\pm 9.2$ & $-0.9\pm 0.5$&60 - 240    \\ 
0.80 & $A_{1g}$ & $4886 \pm 261$ & $-5.1 \pm 3.3$ & $-2.4 \pm 0.9$&70 - 235 \\ 
     & $B_{2g}$ & $1923 \pm 143$  & $-24.1 \pm 5.5$ & $-2.7 \pm 0.5$&70 - 235   \\ 
0.95 & $A_{1g}$ & $3420\pm 146$ & $-3.5 \pm 2.4$ & $-5.9 \pm 0.6 $&60 - 258  \\ 
     & $B_{2g}$ & $1600 \pm 75$ & $-63.1 \pm 3.4$ & $-2.7 \pm 0.2 $&60 - 258 \\  
\end{tabular}
\label{tab:Fitparameter}
\end{ruledtabular}
\end{table*}

\begin{figure*}
    \centering

    \begin{minipage}[t]{0.49\textwidth}
        \centering
        \begin{overpic}[height=5.5cm]{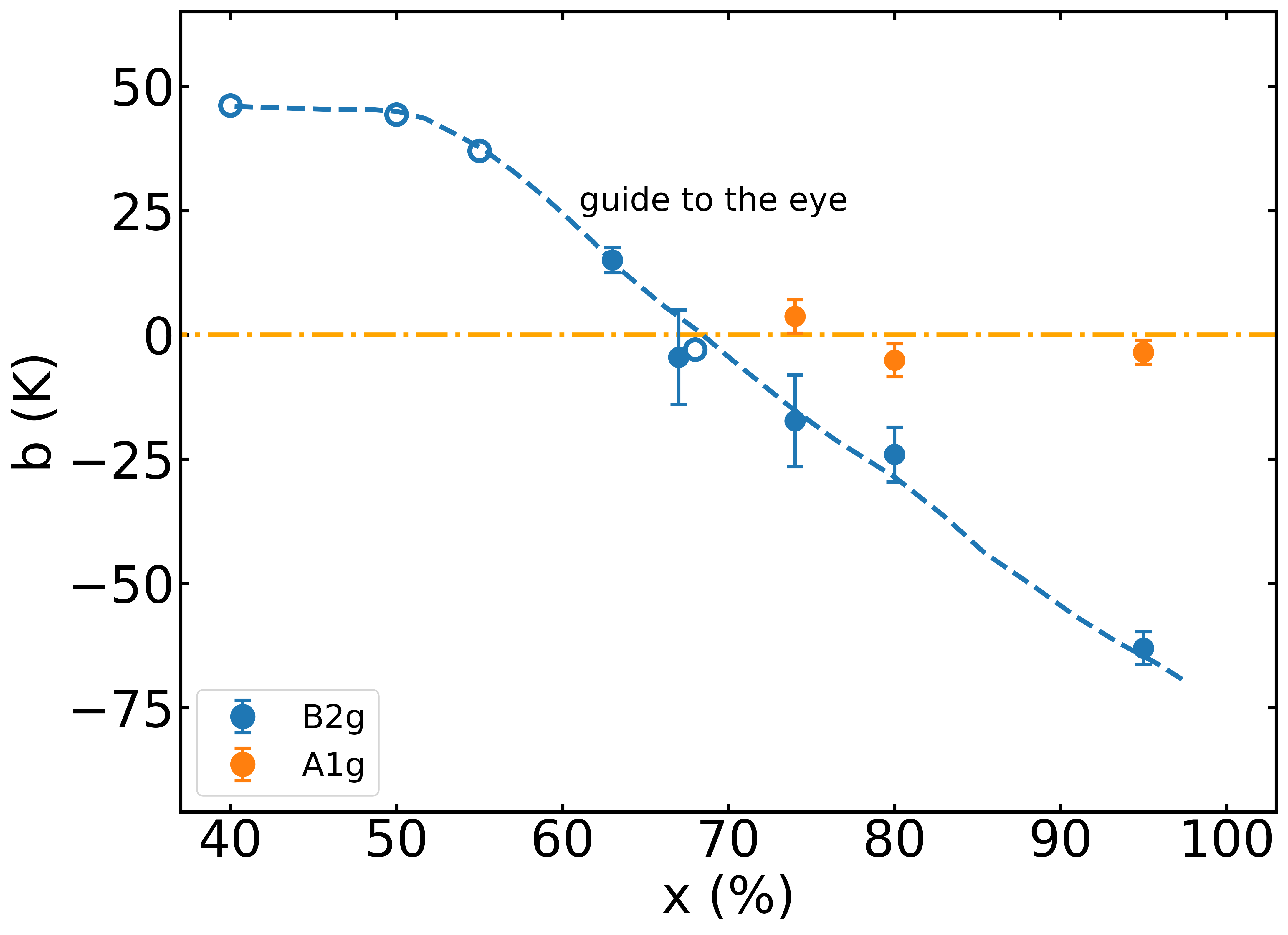}
            \put(2,75){\bfseries (a)}
        \end{overpic}
        \label{fig:Tnem_alles_a}
    \end{minipage}
    \hfill
    \begin{minipage}[t]{0.49\textwidth}
        \centering
        \begin{overpic}[height=5.5cm]{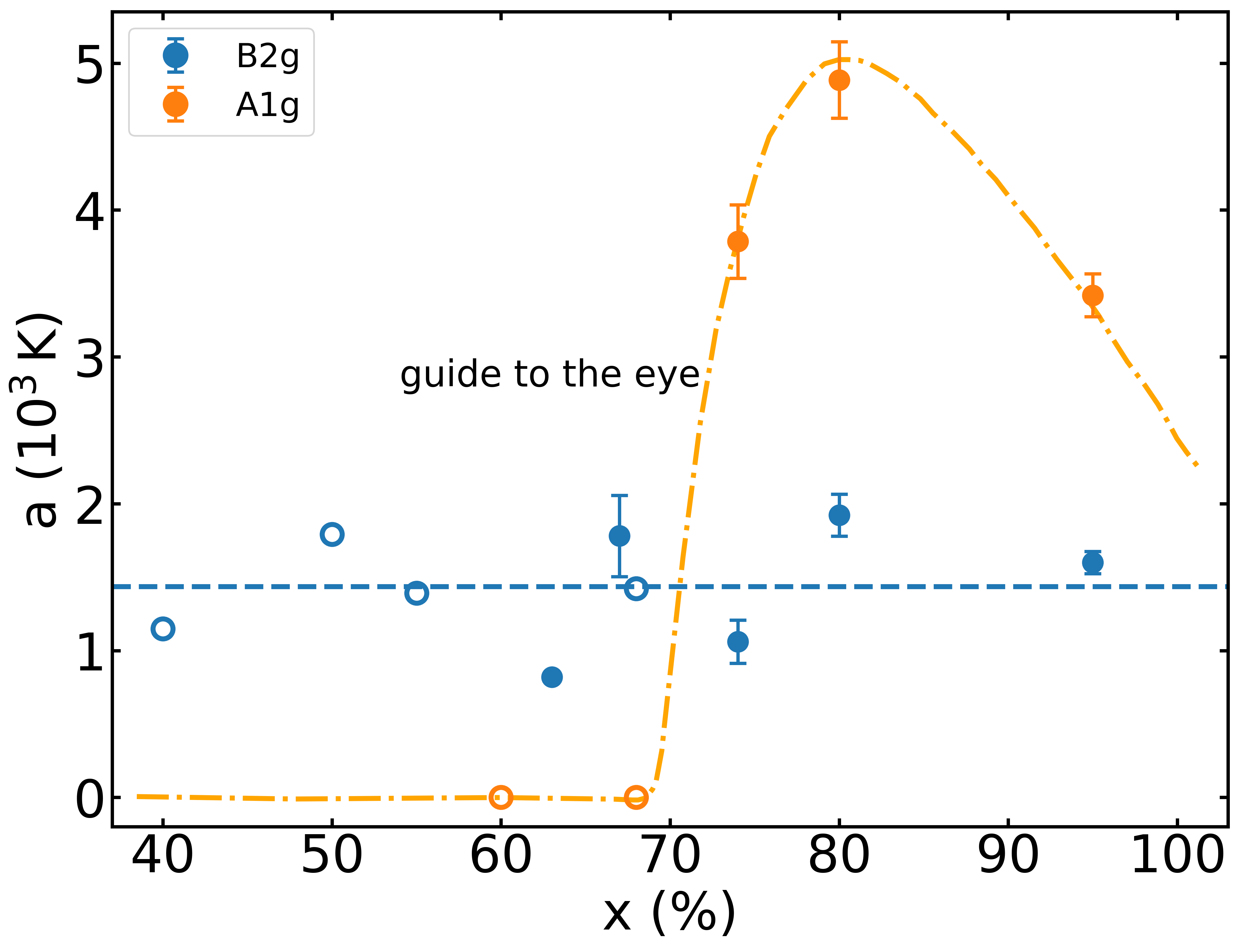}
            \put(0,79){\bfseries (b)}
        \end{overpic}
        \label{fig:Tnem_alles_b}
    \end{minipage}

    \caption{\justifying
    Summary of the Curie-Weiss fit parameters obtained from the separated $A_{1g}$ and $B_{2g}$ channel signals. Open symbols represent literature data and correspond to those shown in Fig. \ref{fig:Ergebnisse}. In (a), the doping dependence of the $b$ parameter is shown, which can be interpreted as the nematic transition temperature $T^{\text{nem}}$ in the $B_{2g}$ channel. Panel (b) displays the doping dependence of the $a$ parameter. The connecting lines between the data points are guides to the eye.}
    \label{fig:Tnem_alles}
\end{figure*}

In Fig. \ref{fig:Tnem_alles}, the Curie-Weiss fit parameters $a$ and $b$ are shown as a function of doping for both symmetry channels. Literature data on elastoresistance measurements in Ba$_{1-x}$K$_x$Fe$_2$As$_2$ with $x = 0.6$ and $x = 0.68$ have shown that the elastoresistive response in these samples consists exclusively of a contribution in the $B_{2g}$ channel~\cite{Wiecki2021,Xiaochen1}, from which it is inferred that this behavior applies to all samples with a doping level below $x = 0.68$. The plot is supplemented by literature data (open symbols) and by the results of Curie-Weiss fits to the longitudinal elastoresistance measurements from Fig. \ref{fig:Ergebnisse} for samples with $x < 0.68$, since these can be attributed to the $B_{2g}$ channel.\\
The fit parameter $b$ remains constant at approximately 45\,K in the $B_{2g}$ channel up to $x = 0.5$, and then decreases nearly linearly to $-63\,\text{K}$ at $x = 0.95$. The zero crossing occurs at $x \approx 0.68$. The three values determined for the $A_{1g}$ channel fluctuate slightly around 0\,K. \\
The $a$ parameter in the $B_{2g}$ channel remains approximately constant for all samples, with an average value of about 1440\,K. For the $A_{1g}$ channel, the parameter is taken to be zero for all samples with $x < 0.68$, since no significant $A_{1g}$ contribution is assumed in these cases. Above this doping level, the parameter increases sharply, reaching a maximum value of approximately 4900\,K at $x \approx 0.8$, and then decreases again to around 3400\,K at $x = 0.95$. \\
The elastoresistance in the $B_{2g}$ channel is proportional to the nematic susceptibility, since the structural distortion associated with the nematic phase transforms according to this symmetry channel \cite{Daten1}. Nematic fluctuations of this kind have been detected via elastoresistance measurements in a wide range of electron-doped iron-based superconductors, such as LaFe$_{1-x}$Co$_x$AsO ($x = 0$–$0.075$) and Ba(Fe$_{1-x}$Co$_x$)$_2$As$_2$ ($x = 0$–$0.14$) \cite{Xiaochen2020LaFe,elektronen1,FeSeS}. In this context, the $b$ parameter extracted from the Curie-Weiss fits is interpreted as the nematic transition temperature $T^{\text{nem}}$ within mean-field theory, indicating the onset of long-range nematic order \cite{Xiaochen2020LaFe}.
For all compounds studied in the literature, $T^{\text{nem}}$ decreases with increasing doping, and the zero crossing typically occurs in the doping range where the superconducting transition temperature reaches its maximum. Furthermore, an enhancement of the $a$ parameter, reflecting the strength of the nematic susceptibility, has been observed in this region. These findings have been interpreted in terms of a nematic quantum critical point \cite{Böhmer2022nematic}.\\
The divergent $B_{2g}$-channel signals measured in this work (hole-doped side of the phase diagram) can likewise be attributed to nematic fluctuations, which justifies interpreting the $b$ parameter of the Curie-Weiss fit as $T^{\text{nem}}$. The observed zero crossing at $x \approx 0.68$ and the pronounced decrease to $-63\,\text{K}$ at $x = 0.95$ indicate that the system exhibits no tendency toward nematic order beyond $x \approx 0.68$, and that increasing doping further suppresses nematic fluctuations. In addition, the $a$ parameter remains constant across the entire doping range at a value of approximately 1450\,K significantly smaller than, for instance, in LaFe$_{1-x}$Co$_x$AsO, where values exceeding $10^4$\,K have been reported \cite{Xiaochen2020LaFe}.\\
The emergence of a measurable divergent elastoresistance in the $A_{1g}$ channel provides clear evidence of strong electronic correlations in heavily hole-doped Ba$_{1-x}$K$_x$Fe$_2$As$_2$ and their sensitivity to symmetric $A_{1g}$ strain \cite{A1gQuelle1288,Wiecki2021,Wieki2020}. Notably, the $a$ parameter which reflects the coupling strength to the electronic correlations exhibits a maximum at about $x=0.8$, too. Hence, the  peak of the global elasotresistance signal at around this doping level shown in Fig. \ref{fig:Ergebnisse}a) may be attributed to such a non-monotonic doping evolution of the electronic correlations, in addition to recently suggested effects of a Lifshitz transition~\cite{Xiaochen1}.

\section{Theoretical modeling}
In order to model the orbital selective electronic correlations in Ba$_{1-x}$K$_x$Fe$_2$As$_2$ and to disentangle the correlation effects from more conventional band effects in the elastotransport, we employ a five-orbital tight-binding Hamiltonian $\mathcal{H} = \mathcal{H}_0 + \mathcal{H}_1$ with local interactions, as introduced in Ref.~\cite{TheorieMott}.

The kinetic part
\begin{align}
\mathcal{H}_0
  &= \sum_{i\neq j} \sum_{mm',\sigma} 
     t^{mm'}_{ij} \, d^\dagger_{im\sigma} d_{jm'\sigma}  \notag \\
  &\quad + \sum_{im\sigma} (\varepsilon_m - \mu)\,
     d^\dagger_{im\sigma} d_{im\sigma}
     \label{H_0}
\end{align}
describes the hopping between Fe sites and the orbital-dependent on-site energies $\varepsilon_m$.  
Here, $t^{mm'}_{ij}$ is the transfer matrix element, which specifies the hopping amplitude from orbital $m$ at site $j$ to orbital $m'$ at site $i$. The chemical potential $\mu$ controls the doping level.  

The interaction term reads:
\begin{align}
\mathcal{H}_1
  &= U \sum_{i,m} n_{im\uparrow} n_{im\downarrow} \notag \\
  &\quad + U' \sum_{i,m>m',\sigma} n_{im\sigma} n_{im'\bar{\sigma}} \notag \\
  &\quad + (U' - J) \sum_{i,m>m',\sigma} n_{im\sigma} n_{im'\sigma}
  \label{H_1}
\end{align}
where $U$ and $U' = U - 2J$ are the intra- and interorbital Coulomb repulsions, respectively, and $J$ is the Hund’s coupling. Here $n_{im\sigma} = d^\dagger_{im\sigma} d_{im\sigma}$ and $\bar{\sigma}$ denotes the opposite spin to $\sigma$.

The interaction part \eqref{H_1} prevents an exact solution of the model. In order to solve the model approximately, but to maintain the low-energy properties of the model, a renormalization method developed in Ref.~\cite{Sykora2020} was used. In this method, the high-energy parts of the model Hamiltonian are eliminated with the help of a unitary transformation and in this way the original Hamiltonian is mapped to an effective low-energy model with renormalized parameters. This new effective model has the same eigenvalues as the original model, but contains only the low-energy interaction processes, which should describe the behavior of the model at low temperatures. The effective model reads
\begin{align}
\mathcal{\tilde{H}}
  &= \sum_{i,j}\sum_{mm',\sigma} 
     \tilde{t}_{ij}^{mm'} d_{im\sigma}^\dagger d_{jm'\sigma} \notag \\
  &\quad - \sum_{i,m>m',\sigma} 
     \tilde{g}_{mm'} \bigl(n_{im\sigma} - n_{im'\sigma}\bigr)^2 \notag \\
  &\quad - \sum_{i,m>m'}  
     \tilde{j}_{mm'} \, {\bf s}_{im} \cdot {\bf s}_{im'}
     \label{eqModell}
\end{align}
All tilde parameters in this equation are connected to the original parameters in equations \eqref{H_0} and \eqref{H_1} via a system of differential equations. These equations are solved numerically following the formalism developed in Ref.~\cite{Sykora2020}. The first term in equation \eqref{eqModell} is an effective hopping and the following two terms describe the relevant low-energy excitations of the original model. 
The coupling parameter $\tilde{g}_{mm'}$ of the first interaction term represents the nematic density-density interaction between orbitals $m$ and $m'$. The second term with the coupling $\tilde{j}_{mm'}$ denotes the spin-spin exchange coupling between different orbitals. Both parameters arise naturally from the original electron-electron interactions through the renormalization procedure. Note that the Hamiltonian \eqref{eqModell} is still not diagonal. It can be diagonalized using another unitary transformation within the same renormalization method and factorization approximations. After this final step any measurable quantity can be calculated.

\begin{figure*}
    \centering

    \begin{minipage}[t]{0.49\textwidth}
        \centering
        \begin{overpic}[height=6.0cm]{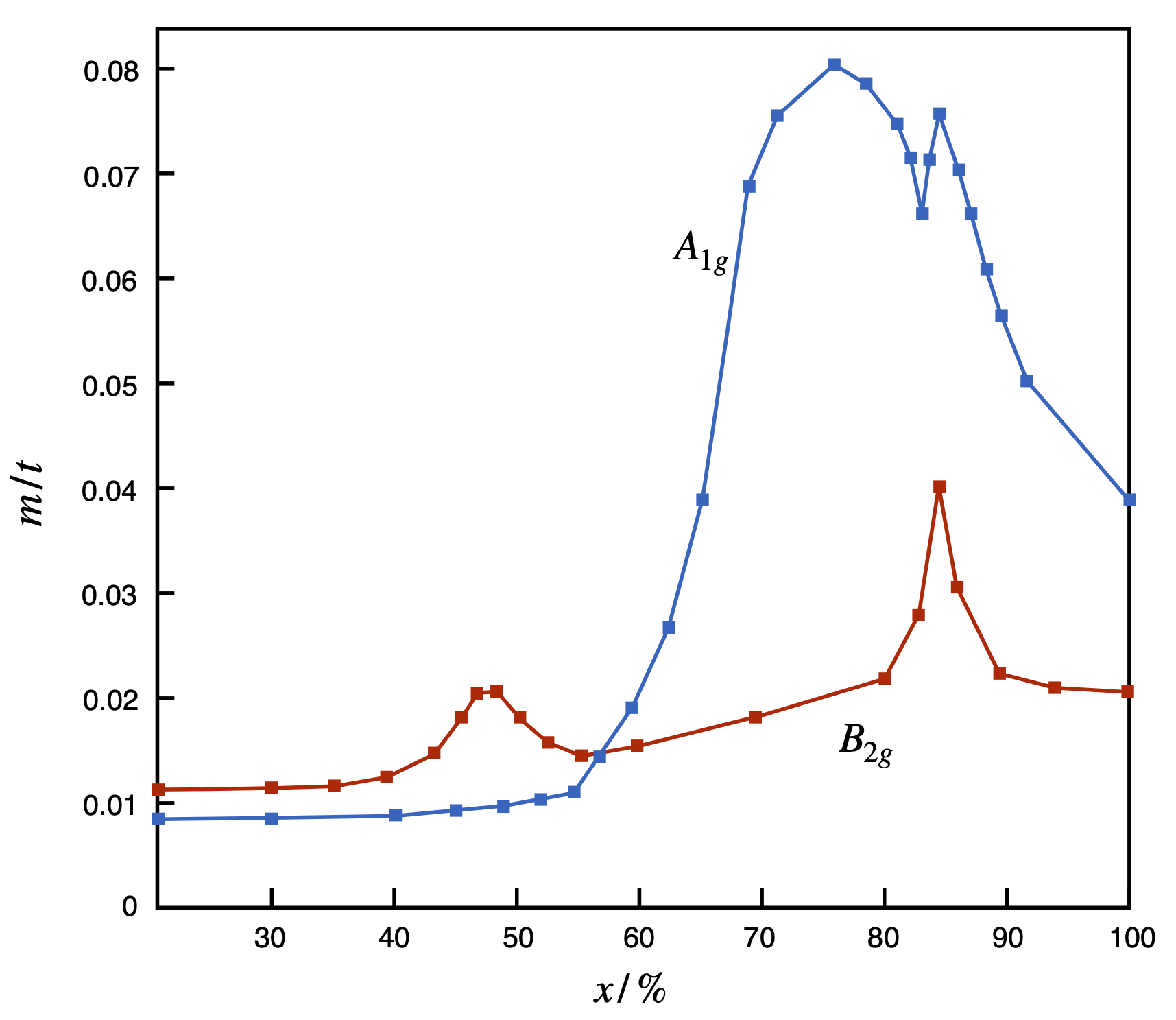}
            \put(2,85){\bfseries (a)}
        \end{overpic}
    \end{minipage}
    \hfill
    \begin{minipage}[t]{0.49\textwidth}
        \centering
        \begin{overpic}[height=6.0cm]{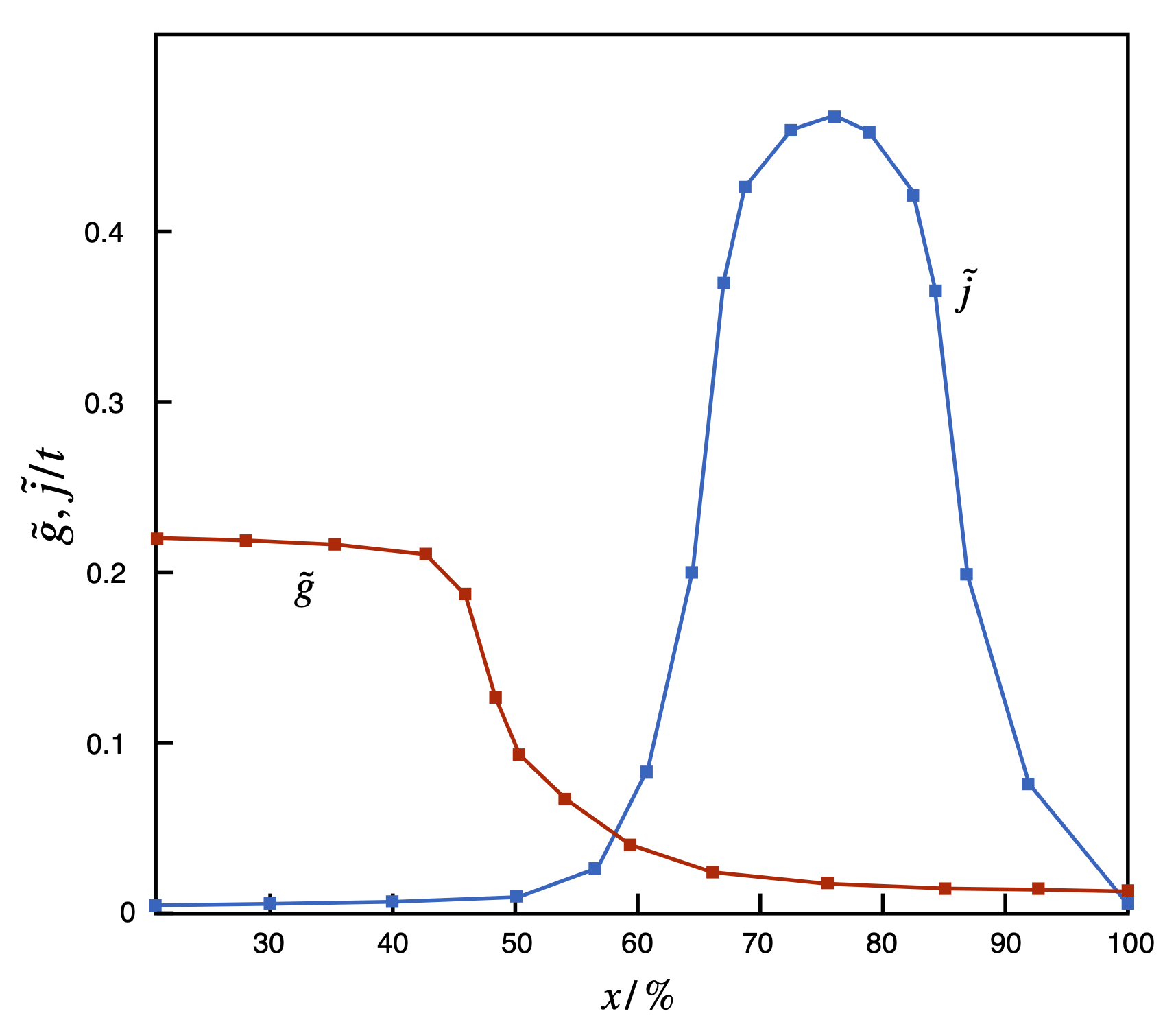}
            \put(2,85){\bfseries (b)}
        \end{overpic}
        
    \end{minipage}

    \caption{\justifying
    (a) Calculated elastoresistance signal of the $A_{1g}$ and $B_{2g}$ channels as defined in equation~\eqref{n_AB}. 
    (b) Calculated values of the effective interactions $\tilde{g}$ (nematic density–density interaction) and $\tilde{j}$ (spin–spin interaction) as a function of doping. 
    The parameter $\tilde{g}$ is predominantly influenced by the $d_{xz/yz}$ orbitals, whereas $\tilde{j}$ is mainly determined by the $d_{xy}$ orbital.}
    \label{fig:Ergebnisse_Berechnung}
\end{figure*}

The described procedure was used to calculate the elastoresistive response via the Kubo-Greenwood approach. In this approach the electrical conductivity was represented as a current-current correlation function and the corresponding expectation value was calculated numerically with the described remormation method.

The effect of applied strain was introduced by modifying the hopping amplitudes \( t_{ij}^{mm'} \) in the tight-binding part of the Hamiltonian, which simulates the deformation of the crystal lattice. In this way, strain-induced changes in the electronic structure were taken into account on a microscopic level.

The conductivity response to strain was expressed in terms of generalized susceptibilities evaluated using the Mori scalar product and a projection formalism. This method allows for a systematic treatment of the relevant fluctuation channels, taking into account the interplay between electronic structure and correlation effects. Two central response coefficients were computed, describing the change in resistivity in longitudinal and transverse directions under small variations of the hopping along the strain axis \( t_x \):
\begin{equation}
n_{xx} = \frac{1}{\sigma^{-1}_{xx}} \lim_{\Delta t_x \to 0} 
\frac{\sigma^{-1}_{xx}(t_x + \Delta t_x) - \sigma^{-1}_{xx}}{\Delta t_x}
\end{equation}

\begin{equation}
n_{xy} = \frac{1}{\sigma^{-1}_{yy}} \lim_{\Delta t_x \to 0} 
\frac{\sigma^{-1}_{yy}(t_x + \Delta t_x) - \sigma^{-1}_{yy}}{\Delta t_x}
\end{equation}

From these two matrix elements, the symmetry-resolved components were obtained via:
\begin{equation}
\label{n_AB}
    n_{A_{1g}} = \frac{1}{2} (n_{xx} + n_{xy}), \quad n_{B_{2g}} = \frac{1}{2} (n_{xx} - n_{xy}),
\end{equation}
corresponding to the $A_{1g}$ and $B_{2g}$ strain channels of the tetragonal point group $D_{4h}$. 

In Fig. \ref{fig:Ergebnisse_Berechnung}a), the calculated contributions $n_{A_{1g}}$ and $n_{B_{2g}}$ of the $A_{1g}$ and $B_{2g}$ channels are shown as a function of doping. They are in good agreement with the experimentally determined values in Fig. \ref{fig:Tnem_alles}b). The $A_{1g}$ channel exhibits a pronounced increase with hole doping, reaching a clear maximum at $x \approx 0.8$. Additionally, a subtle anomaly is observed near $x \approx 0.85$. In contrast, the $B_{2g}$ contribution shows a slight increase with doping, exhibiting weak maxima near $x \approx 0.5$ and $x = 0.85$. Both the anomaly in the $A_{1g}$ channel and the second maximum in the $B_{2g}$ channel can be attributed to a Lifshitz transition, which modifies the Fermi surface topology and influences the electronic response in both symmetry channels.

Figure \ref{fig:Ergebnisse_Berechnung}b) shows the calculated low-energy interactions $\tilde{g} = \sum_{m,m'} \tilde{g}_{mm'}$ and $\tilde{j} = \sum_{m,m'} \tilde{j}_{mm'}$ from the effective model \eqref{eqModell} as a function of doping. It can be seen that for low doping levels ($x \lesssim 0.6$) the system is predominantly governed by the nematic interaction parameter $\tilde{g}$, while for higher doping levels ($x > 0.6$) the spin-spin interaction parameter $\tilde{j}$ becomes dominant. The parameter $\tilde{g}$ represents the nematic density-density interaction and is directly linked to the tendency of the system to develop nematic order. This decrease coincides with the first maximum in the $B_{2g}$ channel, which could be related to enhanced quantum fluctuations. An orbital-resolved analysis reveals that $\tilde{g}$ is nonzero only for contributions involving the $d_{xz}$ and $d_{yz}$ orbitals, which is consistent with the established understanding that nematicity in iron-based superconductors is driven by unequal occupation of these orbitals.

In contrast, the large maximum of the $A_{1g}$ channel at high dopings originates from the spin-spin interaction parameter $\tilde{j}$, whose corresponding term in the effective Hamiltonian describes the effective exchange between spins in orbital $m$ and $m'$. This is illustrated by the blue curve in Fig. \ref{fig:Ergebnisse_Berechnung}b). Furthermore, an investigation of the orbital dependence of $\tilde{j}$ shows that it yields a significant contribution only when considering the interaction between electrons and those in the $d_{xy}$ orbital. This observation is consistent with the concept of orbital-selective Mottness, which predicts that the $d_{xy}$ orbital is the most strongly correlated among the Fe $3d$ orbitals.

\section{Discussion}  

The measurements conducted as part of this work successfully reproduce the increase in elastoresistance of Ba$_{1-x}$K$_x$Fe$_2$As$_2$ around $x \approx 0.8$ reported by Hong et al.~\cite{Xiaochen1}. In combination with the decomposition of the elastoresistance signal into the $A_{1g}$ and $B_{2g}$ symmetry channels, as well as the accompanying theoretical modeling, it becomes clear that this increase cannot be attributed solely to a Lifshitz transition, as previously suggested~\cite{Xiaochen1}. Instead, the rise in elastoresistance can be primarily explained by the emergence of electronic correlations at higher doping levels, which manifest experimentally as an enhancement of the signal in the $A_{1g}$ channel. The theoretical model nevertheless indicates that the Lifshitz transition does have an influence, producing a narrow anomaly in the $A_{1g}$ channel and a corresponding local maximum in the $B_{2g}$ channel near $x \approx 0.85$.
However, due to the limited doping resolution of the measurements and the sharpness of these features, they cannot be clearly resolved in the experimental data. Furthermore, a divergent signal in the $B_{2g}$ channel was detected throughout the entire doping range considered, which can be interpreted as a remnant of nematic fluctuations.

Our interpretation of the orbital sensitive transport measurements indicate that a band with predominant $d_{xy}$ character strongly interacts with electrons in the remaining orbitals, giving rise to quasi-localized $d_{xy}$ electrons consistent with the orbital-selective Mott scenario~\cite{TheorieMott}. According to our effective Hamiltonian (Eq. \ref{eqModell}) this interaction is of the nature of an exchange coupling between localized spins in the $d_{xy}$ orbital and itinerant electrons, reflecting the characteristics of a Kondo effect. In this way localized spins can act as magnetic scattering centers for the mobile electrons in the other, more delocalized orbitals. In line with this picture, Corbae et al.~\cite{Elena} have shown by ARPES that in hole-doped Ba$_{1-x}$K$_x$Fe$_2$As$_2$ the superconducting gap on the \textit{d}$_{xy}$ orbital indeed vanishes already at intermediate doping, while the gaps on the more itinerant \textit{d}$_{xz/yz}$ orbitals persist. This provides direct spectroscopic evidence that the localized nature of the \textit{d}$_{xy}$ states suppresses superconductivity in this orbital, consistent with our interpretation of quasi-localized \textit{d}$_{xy}$ electrons acting as magnetic scattering centers.

\begin{figure}
    \centering
    \includegraphics[width=0.48\textwidth]{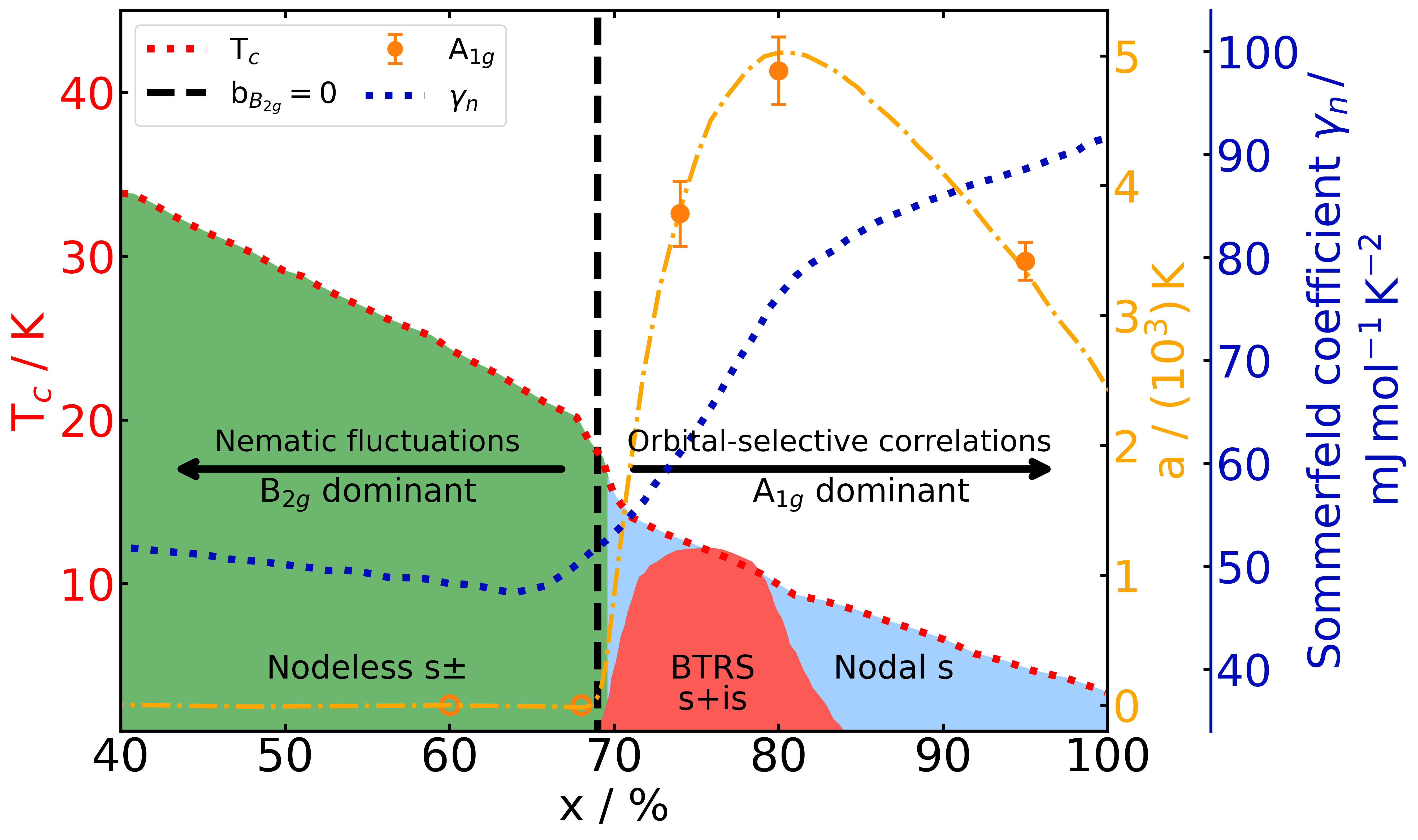} 
    \caption{Phase diagram from Ref. \cite{Vadim2020}, supplemented with the zero crossing of the $b$ parameter and the $a$ parameter of the $A_{1g}$ channel. } 
    \label{fig:Phasendiagramm_alles} 
\end{figure}

In Fig. \ref{fig:Phasendiagramm_alles}, the phase diagram from Ref.~\cite{Vadim2020} is extended by including the zero crossing of the $b$ parameter in the $B_{2g}$ channel (indicated by the vertical black line) and the doping dependence of the $a$ parameter in the $A_{1g}$ channel (yellow dashed line). Both parameters appear to be potentially connected to the superconducting mechanism. The zero crossing of $b$ occurs at approximately the same doping level at which a step-like change in the superconducting transition temperature $T_c$ is observed and where the BTRS phase is believed to begin. Simultaneously, the $A_{1g}$ signal becomes measurable from approximately the same doping level onward.  
Another indication of the involvement of correlations in the increased response in the $A_{1g}$ channel is the course of the Sommerfeld coefficient as a function of doping (represented by a blue dotted line in Fig. \ref{fig:Phasendiagramm_alles}). This quantity is a measure of an increased density of states at the Fermi energy and represents an increased effective mass. It can be seen that a particularly enhanced value of the Sommerfeld coefficient sets in at about the same doping as the measured transport in the $A_{1g}$ channel as well as also the BTRS phase (red area). This indicates that the superconducting order parameter of the BTRS phase could be mediated by strongly orbital selective correlations. All this indicates that for $x > 0.68$ qualitatively new physics emerges. In particular, our findings suggest that the origin of the BTRS superconducting phase is strongly connected to the orbital-selective correlations, while the evidence of weak electronic correlations nevertheless shows that these also still play a role. Due to the limited number of data points, however, no definitive conclusion can be drawn regarding the interplay between the $A_{1g}$ response and the onset of the BTRS phase. 

Ishida et al.~\cite{LochdotiertandereMessung} carried out elastoresistance measurements on a series of Ba$_{1-x}$Rb$_x$Fe$_2$As$_2$ single crystals along both the [110] and [100] crystallographic directions. At high doping levels, they observed a shift of the divergent elastoresistance from the [110] to the [100] direction, which they interpreted as a signature of XY-nematic fluctuations. However, no symmetry decomposition of the measured signals was performed. Moreover, their measurements along the [110] direction also revealed a minimum in the $b$ parameter at $x \approx 0.75$. This similarity suggests that, as in Ba$_{1-x}$K$_x$Fe$_2$As$_2$, the divergent elastoresistance in Ba$_{1-x}$Rb$_x$Fe$_2$As$_2$ may also be dominated by a contribution from the $A_{1g}$ channel, an interpretation that was not examined further in the study by Ishida et al.~and has also been discussed by Wiecki et al. \cite{Wiecki2021}.

\section{Summary}
In this work, nematic fluctuations and electronic correlations in the hole-doped iron pnictide superconductor Ba$_{1-x}$K$_x$Fe$_2$As$_2$ were investigated using longitudinal and transverse elastoresistance measurements which were analyzed by a many-particle modeling. The decomposition into $A_{1g}$ and $B_{2g}$ channels reveals a doping-dependent change in the dominant contribution, with nematic fluctuations in the $B_{2g}$ channel prevailing at low doping and the $A_{1g}$ channel dominating at high doping, where it exhibits a pronounced maximum at $x \approx 0.8$.
We attribute the $A_{1g}$ enhancement to a selective interaction of the $d_{xy}$ orbital to the other itinerant electrons in form of an effective spin-spin interaction. Despite the dominance of the $A_{1g}$ signal, weakened nematic fluctuations in the $B_{2g}$ channel remain detectable. The results demonstrate that electronic correlations in the strongly hole-doped regime have a significant impact on superconductivity and highlight the value of elastoresistance measurements for investigating the interplay between nematic and correlation-driven physics in iron-based superconductors.

\begin{acknowledgments}
We thank Tino Schreiner and Dany Baumann for their technical support. We are also grateful to Elena Corbae, Zhi-Xun Shen, Donghui Lu, Rong Zhang, and Anna Böhmer for valuable discussions. S.S. acknowledges funding from the Deutsche Forschungsgemeinschaft (DFG) via the Cluster of Excellence on Complexity and Topology in Quantum Matter ct.qmat (EXC 2147, project id 390858490).
V.G. is supported by the NSFC grant 12374139.

\end{acknowledgments}

\bibliography{literatur}

\end{document}